\documentclass{svmult}
\usepackage{amssymb}
\usepackage{graphicx}    % standard LaTeX graphics tool
\def\beq{\begin{equation}}
\def\eeq{\end{equation}}

\def\bea{\begin{eqnarray}}
\def\eea{\end{eqnarray}}
\def\ba{\begin{array}}                  %array
\def\ea{\end{array}}

\def\g{\gamma}

\begin{document}

\title*{Quantum Gravity Phenomenology\\ and Lorentz 
Violation\thanks{Expanded 
version of a lecture by T. Jacobson,
to be published in {\em Particle Physics
and the Universe, Proceedings of the 9th Adriatic Meeting}, eds.
J.~Trampetic and J.~Wess (Springer-Verlag, 2004)}} 
\author{Ted Jacobson\inst{1},
Stefano Liberati\inst{2}\and
David Mattingly\inst{3}}
% Please full name, not just initials!
\institute{Institut d'Astrophysique de Paris, 98bis bd Arago, 
75014 Paris France, and \\
Department of Physics,
University of Maryland, College Park, MD 20742 USA
\texttt{jacobson@umd.edu}
\and SISSA, Via Beirut 2-4, 34014 Trieste, Italy and INFN Trieste
\texttt{liberati@sissa.it}
\and Department of Physics, University of California at Davis
\texttt{mattingly@physics.ucdavis.edu}}%
\maketitle

In
most fields of physics it goes without saying that
observation and prediction play a central role, but
unfortunately quantum gravity (QG) has so far not fit
that mold.  Many intriguing and ingenious ideas have been
explored, but it seems safe to say that without both
observing phenomena that depend on QG, and extracting
reliable predictions from candidate theories that can be
compared with observations, the goal of a theory capable
of incorporating quantum mechanics and general relativity
will remain unattainable.

Besides the classical limit, there is one observed phenomenon for
which quantum gravity makes a prediction that has received
encouraging support: the scale invariant spectrum of primordial
cosmological perturbations. The quantized longitudinal linearized
gravitational mode, albeit slave to the inflaton and not a
dynamically independent degree of freedom, plays an essential role
in this story~\cite{Mukhanov:1990me}.

What other types of phenomena might be characteristic of
a quantum gravity theory?  Motivated by tentative
theories, partial calculations, intimations of symmetry
violation, hunches, philosophy, etc, some of the proposed
ideas are: loss of quantum coherence or state collapse,
QG imprint on initial cosmological perturbations, scalar
moduli or other new fields, extra dimensions and
low-scale QG, deviations from Newton's law, black holes
produced in colliders, violation of global internal
symmetries, and violation of spacetime symmetries. It is
this last item, more specifically the possibility of
Lorentz violation (LV), that is the focus of this
article.

From the observational point of view, technological developments
are encouraging a new look at the possibility of LV.  Increased
detector size, space-borne instruments, technological improvement,
and technique refinement are permitting observations to probe
higher energies, weaker interactions, lower fluxes, lower
temperatures, shorter time resolution, and longer distances.  It
comes as a welcome surprise that the day of true quantum gravity
observations may not be so far off~\cite{Dawn}

%%%%%%%%%%%%%%%%%%%%%%%%%%%%%%%%%%%%
\section{Lorentz violation?}
\label{sec:LV?}
%%%%%%%%%%%%%%%%%%%%%%%%%%%%%%%%%%%%

Lorentz symmetry is linked to a scale-free nature of
spacetime: unbounded boosts expose ultra-short distances,
and yet nothing changes.  However, suggestions for
Lorentz violation have
come from: the need to cut off UV
divergences of quantum field theory and of black hole
entropy, tentative calculations in various QG scenarios
(e.g. semiclassical spin-network calculations in Loop QG, string
theory tensor VEVs, non-commutative geometry, some
brane-world backgrounds), and the possibly missing GZK
cutoff on ultra high energy (UHE) cosmic rays.

The GZK question has generated a lot of interest, and is
to date the only phenomenon that might point to a
breakdown of standard physics in quantum gravity, hence
we take a moment to discuss it. The idea~\cite{GZK} is
that in collisions of ultra high energy protons with
cosmic microwave background photons there can be
sufficient energy in the center of mass frame to create a
pion, leading to the the reaction
\begin{equation}
p+\g_{\rm CMB}\rightarrow p+\pi.
\end{equation}
In this way, the initial proton energy is degraded with
an attenuation length of about 50 Mpc.  Since plausible
astrophysical sources for UHE particles are
located at distances larger than 50 Mpc, one expects
a cutoff in the
cosmic ray proton energy spectrum at around
$5\times10^{19}$ eV for protons coming from beyond a few
megaparsecs.  If Lorentz symmetry is violated, then the
energy threshold for this reaction could be lowered,
raised, or removed entirely, or an upper threshold where
the reaction cuts off could even be introduced
(see e.g.~\cite{Jacobson:2002hd} and references therein).

One of the experiments measuring the UHE cosmic ray
spectrum, the AGASA experiment, has not seen the
cutoff. An analysis~\cite{DeMarco:2003ig} from January
2003 concluded that the cutoff was absent at the 2.5
sigma level, while another experiment, Hi-Res, is
consistent with the cutoff but at a lower confidence
level.  The question should be answered in the near
future by the AUGER observatory, a combined array of 1600
water \v{C}erenkov detectors and 24 telescopic air
flouresence detectors under construction on the Argentine
pampas~\cite{Auger}.  The new observatory will see an
event rate one hundred times higher, with better
systematics.

Trans-GZK cosmic rays are not
the only window of opportunity we have to detect or constrain
Lorentz violation induced by QG effects.
In fact, many phenomena accessible to current
observations/experiments are sensitive to
possible violations of Lorentz
invariance.
A partial list is
\begin{itemize}
\item sidereal variation of LV
couplings as the lab moves with
  respect to a preferred frame or directions
\item long baseline dispersion and vacuum birefringence
(e.g.~of signals from gamma ray bursts, active galactic nuclei,
  pulsars, galaxies)
\item new reaction thresholds (e.g.~photon decay, vacuum \v{C}erenkov effect)
\item shifted thresholds (e.g.~photon annihilation from
  blazars, GZK reaction)
\item maximum velocity (e.g.~synchrotron peak from supernova remnants)
\item dynamical effects of LV background fields (e.g.
  gravitational coupling and additional wave modes)
\end{itemize}

We conclude this section with a brief historical overview
including some of the more influential papers but by no means
complete. Suggestions of possible LV in particle physics go back
at least to the 1960's, when a number of authors wrote on that
idea~\cite{Dirac}\footnote{It is
%, Bjorken, Phillips,
 % Blokhintsev, Pavlopoulos, Kirzhnits}.
 amusing to note that Kirzhnits and Chechin in~\cite{Dirac}
 explore the
  possibility that an apparent missing cutoff in the UHE
  cosmic ray spectrum could be explained by something
  that looks very similar to the recently proposed
  ``doubly special relativity''~\cite{DSR}.}.  The
possibility of LV in a metric theory of gravity was
explored beginning at least as early as the
1970's~\cite{LVmetr}. Such
theoretical ideas were
pursued in the '70's and '80's notably by Nielsen and several
other authors on the particle theory side~\cite{7080theory}, and
by Gasperini~\cite{Gasp} on the gravity side.
A number of observational limits were obtained during this
period~\cite{HauganAndWill}.

Towards the end of the 80's Kostelecky and
Samuel~\cite{KS} presented evidence for possible
spontaneous LV in string theory, and
motivated by this explored
LV effects in gravitation.  The role of
Lorentz invariance in the ``trans-Planckian puzzle" of
black hole redshifts and the Hawking effect was
emphasized in the early 90's~\cite{TedUltra}. This led to study of
the Hawking effect for quantum fields with LV dispersion
relations commenced by Unruh~\cite{Unruh} and followed up by
others. Early in the third millenium this line of
research led to work on the related question of the
possible imprint of trans-Planckian frequencies on the
primordial fluctuation spectrum~\cite{BM}.

Meanwhile the consequences of LV for particle physics
were being explored using LV dispersion relations e.g. by
Gonzalez-Mestres~\cite{GM}, and a systematic extension of
the standard model of particle physics incorporating all
possible LV in the renormalizable sector was developed by
Colladay and Kosteleck\'{y}~\cite{CK}.  This latter work
provided a framework for computing the observable
consequences for any experiment and led to much
experimental work setting limits on the LV parameters in
the lagrangian~\cite{AKbook}. Around the same time
Coleman and Glashow suggested the possibility that LV was
the culprit in the possibly missing GZK
cutoff~\cite{CG-GZK}, and explored many other
consequences of renormalizable, isotropic LV leading to
different limiting speeds for different
particles~\cite{CGlong}.

Also at that time it was pointed out by Amelino-Camelia
et al~\cite{GAC-Nat} that the sharp high energy signals
of gamma ray bursts could reveal LV photon dispersion
suppressed by one power of energy over the mass $M\sim
10^{-3}M_{\rm P}$, tantalizingly close to the Planck
mass.  Shortly afterwards Gambini and Pullin~\cite{GP}
argued that semiclassical loop quantum gravity suggests
just such LV.
(Some later work supported this notion, but a recent paper by
Kozameh and Parisi~\cite{KP} argues the other way.)  In any case
the theory is not under enough control at this time to make any
definite statements.

A very strong constraint on photon birefringence was
obtained by Gleiser and Kozameh~\cite{GK} using UV light
from distant galaxies, and if the recent measurement of
polarized gamma rays from a GRB hold up to further
scrutiny this constraint will be further strengthened
dramatically~\cite{JLMS,Mitro}.  Further stimulus came from
the suggestion~\cite{PM} that an LV threshold shift might
explain the apparent under-absorption on the cosmic IR
background of TeV gamma rays from the blazar Mkn501,
however it is now believed by many that this anomaly goes
away when a corrected IR background is used~\cite{Kono}.

The extension of the effective field theory framework to
include LV dimension 5 operators was introduced by Myers
and Pospelov~\cite{MP}, and used to strengthen prior
constraints. Also this framework was used to deduce a
very strong constraint~\cite{Crab} on the possibility of a
maximum electron speed less than the speed of light
from observations of synchrotron
radiation from the Crab Nebula.

\section{Theoretical framework for LV}

Various different theoretical approaches to LV have been taken
to further pursue the ideas summarized above.
Some researchers restrict attention to LV
described in the framework of effective field theory (EFT), while others
allow for effects not describable in this way, such as those
that might be due to stochastic fluctuations
of a ``space-time foam''. Some restrict to rotationally
invariant LV, while others consider also rotational
symmetry breaking. Both true LV as well as
``deformed" Lorentz symmetry 
(in the context of so-called ``doubly special
relativity"\cite{DSR}) have been pursued. Another
difference in approaches is whether one allows for
distinct LV parameters for different particle types,
or proposes a more universal form of LV.

The rest of this article will focus on just one of these
approaches, namely LV describable by standard EFT,
assuming rotational invariance, and allowing distinct LV
parameters for different particles.  In exploring the
possible phenomenology of new physics, it seems useful to
retain enough standard physics so that a) clear
predictions can be made, and b) the possibilities are
narrow enough to be meaningfully constrained.

This approach is not universally favored. For example a
sharp critique appears in~\cite{GAC-crit}. Therefore we
think it is important to spell out the motivation for the
choices we have made. First, while of course it may be
that EFT is not adequate for describing the leading
quantum gravity phenomenology effects, it has proven
itself very effective and flexible in the past. It
produces local energy and momentum conservation laws, and
seems to require for its validity just locality and local
spacetime translation invariance above some length
scale. It describes the standard model and general
relativity (which are presumably not fundamental
theories), a myriad of condensed matter systems at
appropriate length and energy scales, and even string
theory (as perhaps most impressively verified in the
calculations of black hole entropy and Hawking radiation
rates).  It is true that, e.g., non-commutative geometry
(NCG) seems to lead to EFT with problematic IR/UV mixing,
however this more likely indicates a physically
unacceptable feature of such NCG rather than a physical
limitation of EFT.

The assumption of rotational invariance is motivated by the idea
that LV may arise in QG from the presence of a short distance
cutoff. This suggests a breaking of boost invariance, with a
preferred rest frame, but not necessarily rotational invariance.
Since a constraint on pure boost violation is, barring a
conspiracy, also a constraint on boost plus rotation violation, it
is sensible to simplify with the assumption of rotation invariance
at this stage.

Finally why do we choose to complicate matters by allowing for
different LV parameters for different particles? First, EFT for
first order Planck suppressed LV (see section 2.1) requires this
for different polarizations or spin states, so it is unavoidable
in that sense. Second, we see no reason a priori to expect these
parameters to coincide.  The term ``equivalence principle'' has
been used to motivate the equality of the parameters. However, in
the presence of LV dispersion relations, particles with different
masses travel on different trajectories even if they have the same
LV parameters~\cite{Fischbach:wq,Jacobson:2002hd}.  Moreover,
different particles would presumably interact differently with the
spacetime microstructure since they interact differently with themselves
and with each other. An example of this occurs in the braneworld model
discussed in Ref.~\cite{Burgess}, and 
an extreme version occurs in the proposal of
Ref.~\cite{emn} in which only certain particles feel the spacetime
foam effects. (Note however that in this proposal the LV
parameters fluctuate even for a given kind of particle, so EFT
would not be a valid description.)

\subsection{Deformed dispersion relations}
\label{subsec:disp}

A simple approach to a phenomenological
description of LV is via deformed dispersion
relations. If rotation invariance
and integer powers of momentum are assumed
in the expansion of $E^2(\vec{p})$,
the dispersion relation for a given particle type
can be written as
\begin{equation}
E^2=p^2 + m^2 + \Delta(p),
\end{equation}
where $p$ is the magnitude of the three-momentum,
and
\beq
\Delta(p)= \tilde{\eta}_1 p^1 + \tilde{\eta}_2 p^2
+ \tilde{\eta}_3 p^3 + \tilde{\eta}_4 p^4 +\cdots
\label{disprel1}
\eeq
Let us introduce two mass scales, $M=10^{19}\, {\rm
  GeV}\approx M_{\rm Planck}$, the putative scale of
quantum gravity, and $\mu$, a particle physics mass
scale.  To keep mass dimensions explicit we factor out
possibly appropriate powers of these scales, defining the
dimensionful $\eta$'s in terms of corresponding
dimensionless parameters. It might seem natural that the
$p^n$ term with $n\ge3$ be suppressed by $1/M^{n-2}$, and
indeed this has been assumed in most work. But following
this pattern one would expect the $n=2$ term to be
unsuppressed and the $n=1$ term to be even more
important. Since any LV at low energies must be small,
such a pattern is
untenable.
Thus either there is a symmetry or some other mechanism
protecting the lower dimension oprators from large LV, or
the suppression of the higher dimension operators is
greater than $1/M^{n-2}$.  This is an important issue to
which we return later in this article.

For the moment we simply follow the observational lead and
insert at least one inverse power of $M$ in each term, viz.
\beq
\tilde{\eta}_1=\eta_1 \frac{\mu^2}{M},\qquad
\tilde{\eta}_2=\eta_2 \frac{\mu}{M},\qquad
\tilde{\eta}_3=\eta_3 \frac{1}{M},\qquad
\tilde{\eta}_4=\eta_4 \frac{1}{M^2}.
\label{disprel2}
\eeq
%
%t30
In characterizing the strength of a constraint
we refer to the $\eta_n$ without the tilde,
so we are comparing to what might be expected from
Planck-suppressed LV.
We allow the LV parameters $\eta_i$ to depend on the
particle type, and indeed it turns out that they {\it
  must} sometimes be different but related in certain
ways for photon polarization states, and for particle and
antiparticle states, if the framework of effective field
theory is adopted. In an even more general setting,
Lehnert~\cite{Leh} studied theoretical constraints on
this type of LV and deduced the necessity of some of
these parameter relations.

This general framework allows for superluminal
propagation, and spacelike 4-momentum relative to a fixed
background metric.  It has been argued~\cite{KL} that this
may lead to problems with causality and stability, but we
do not share this opinion. In the context of a LV theory,
there can be a preferred reference frame.  As long as the
physics is guaranteed to be causal and the states all
have positive energy in the preferred frame, we cannot
see any room for such problems to arise.

\subsection{Effective field theory and LV}

The standard model extension (SME) of Colladay and
Kosteleck\'{y}~\cite{CK} consists of the standard model of
particle physics plus all Lorentz violating renormalizable
operators (i.e. of mass dimension $\le4$) that can be written
without changing the field content or violating the gauge
symmetry. For illustration, the leading order terms in the QED
sector are the dimension three terms
\beq -b_a\bar{\psi}\gamma_5 \gamma^a \psi
-\frac{1}{2}H_{ab}\bar{\psi}\sigma^{ab}\psi \eeq
and the dimension four terms
\beq -\frac{1}{4}k^{abcd}F_{ab}F_{cd}
+\frac{i}{2}\bar{\psi}(c_{ab}+ d_{ab}\gamma_5)\gamma^a
\stackrel{\leftrightarrow}{D^b}\psi, \eeq
where the dimension one coefficients $b_a$, $H_{ab}$ and
dimensionless $k^{abcd}$, $c_{ab}$, and $d_{ab}$ are constant
tensors characterizing the LV.  If we assume rotational invariance
then these must all be constructed from a given unit timelike
vector $u^a$ and the Minkowski metric $\eta_{ab}$, hence
$b_a\propto u_a$, $H_{ab}=0$, $k^{abcd}\propto u^{[a} \eta^{b][c}
  u^{d]}$, $c_{ab}$ and $d_{ab}\propto u_au_b$.
Such LV is thus characterized by just four numbers.

The study of Lorentz violating EFT in the higher mass dimension
sector was initiated by Myers and Pospelov~\cite{MP}.  They
classified all LV dimension five operators that can be added to
the QED Lagrangian and are quadratic in the fields, rotation
invariant, gauge invariant, not
reducible to lower
and/or higher dimension operators using the field equations, and
contribute $p^3$ terms to the dispersion relation. Again, just
three parameters arise:
\beq
\frac{\xi}{M}u^mF_{ma}(u\cdot\partial)(u_n\tilde{F}^{na})+\frac{1}{M}
u^m\bar{\psi}\gamma_m(\zeta_1+\zeta_2\gamma_5)(u\cdot
\partial)^2\psi \label{dim5} \eeq
where $\tilde{F}$ denotes the dual of $F$. All of these terms
violate CPT symmetry as well as Lorentz invariance. Thus if one
knew CPT were preserved, these LV operators would be
forbidden.

In the limit of high energy $E\gg m$, the photon and
electron dispersion relations following from QED with the
above terms are~\cite{MP,JLMS}
\bea
\omega_{R,L}^2&=& k^2 \pm \frac{2\xi}{M}k^3\\
E_{\pm}^2&=& p^2 + m^2  +\frac{2(\zeta_1\pm\zeta_2)}{M}p^3.
\label{QEDdisp}
\eea
The photon subscripts $R$ and $L$ refer to right and left
circular polarization, hence these necessarily have
opposite LV parameters. The electron subscripts $\pm$
refer to the helicity, which can be shown to be a good
quantum number in the presence of these LV
terms~\cite{JLMS}.  Moreover, if we write
$\eta_\pm=2(\zeta_1\pm\zeta_2)$ for the LV parameters of
the two electron helicities, those for positrons 
% of positive and negative helicity 
are given by~\cite{JLMS}
\beq
\eta^{\rm positron}_\pm=-\eta^{\rm electron}_\mp.
\eeq
%$-\eta_-$ and $-\eta_+$ respectively 
%This means that
%earlier constraints from photon decay $\gamma\rightarrow
%e^+e^-$ and photon absorption $\gamma\gamma\rightarrow
%e^+e^-$ (which were derived for the special case $\zeta_2=0
%\,\Rightarrow \eta_+=\eta_-$) must be re-analyzed.

\subsection{Un-naturalness of small LV at low energy}

As discussed above in subsection \ref{subsec:disp},
if LV operators of dimension $n>4$ are suppressed,
as we have imagined, by $1/M^{n-2}$, LV would
feed down to the lower dimension operators
and be strong at low
energies~\cite{CGlong,MP,PerezSudarsky,Collins},
unless there is a symmetry or some other mechanism
that protects operators of dimension four and less from
strong LV. What symmetry (other than Lorentz invariance, of course!)
could that possibly be?

In the Euclidean context, a discrete subgroup of the
Euclidean rotation group suffices to protect the
operators of dimension four and less from violation
of rotation symmetry. For example~\cite{Hyper},
consider the ``kinetic'' term in the
EFT for a scalar field with hypercubic symmetry,
$M^{\mu\nu}\partial_\mu\phi\partial_\nu\phi$.
The only tensor $M^{\mu\nu}$ with hypercubic
symmetry is proportional to the Kronecker delta
$\delta^{\mu\nu}$, so full rotational invariance
is an ``accidental'' symmetry of the kinetic operator.

If one tries to mimic this construction on a Minkowski lattice
admitting a discrete subgroup of the Lorentz group, one faces the
problem that each point has an infinite number of neighbors
related by the Lorentz boosts. For the action to share the
discrete symmetry each point would have to appear in infinitely
many terms of the discrete action, presumably rendering the
equations of motion meaningless.

Another symmetry that could do the trick is three dimensional
rotational symmetry together with a symmetry between different
particle types. For example, rotational symmetry would imply that
the kinetic term for a scalar field takes the form
$(\partial_t\phi)^2-c^2(\nabla\phi)^2$, for some constant $c$.
Then for multiple scalar fields, a symmetry relating the fields
would imply that the constant $c$ is the same for all, hence the
kinetic term would be Lorentz invariant with $c$ playing the role
of the speed of light.  Unfortunately this mechanism does not work
in nature, since there is no symmetry relating all the physical
fields.  Perhaps under some conditions a partial symmetry could be
adequate, e.g. grand unified gauge and/or super symmetry.

We are thus in the uncomfortable position of lacking any
theoretical realization of the Lorentz symmetry breaking scheme
upon which constraints are being imposed. This does not mean that
no realization exists, but it is worrisome. If none exists, then
our parametrization (\ref{disprel2}) is misleading, since there
should be more powers of $1/M$ suppressing the higher dimension
terms, likely rendering any constraints on those terms
uninteresting.

\section{Constraints}

Observable effects of LV arise, among other things, from
1)
%t30
%time dependence
sidereal variation
of LV couplings due to motion of the
laboratory relative to the preferred frame, 2) dispersion
and birefringence of signals over long travel times, 3)
anomalous reaction thresholds.  We will often express the
constraints in terms of the dimensionless parameters
$\eta_n$ introduced in (\ref{disprel2}). An order unity
value might be considered to be expected in Planck
suppressed LV.

The possibility of interesting constraints in spite of
Planck suppression arises 
in different ways for the different types of observations.
In the laboratory
experiments looking for sidereal variations, the enormous
number of atoms allow a resonance frequency to be measured
extremely accurately. In the case of
dispersion or birefringence, the enormous propagation
distances would allow a tiny effect to accumulate. In the
anomalous threshold case, the creation of a particle with
mass $m$ would be strongly affected by a LV term when
the momentum becomes large enough for this term to be
comparable to the mass term in the dispersion relation.

Consider first the case $n=2$.  For the $n=2$ term in
(\ref{disprel1},\ref{disprel2}), 
the absence of a strong threshold effect
yields a constraint $\eta_2 \lesssim (m/p)^2(M/\mu)$.  If we
consider protons and put $\mu=m=m_p\sim 1$ GeV, this gives an
order unity constraint when $p\sim \sqrt{mM} \sim 10^{19}$ eV.
Thus the GZK threshold, if confirmed, can give an order unity
constraint, but multi-TeV astrophysics yields much weaker
constraints.  The strongest laboratory constraints on dimension
three and four operators come from clock comparison experiments
using noble gas masers~\cite{Bear:2000cd}.  The constraints limit
a combination of the coefficients for dimension three and four
operators for the neutron to be below $10^{-31}$ GeV (the
dimension four coefficients are weighted by the neutron mass,
yielding a constraint in units of energy).  Astrophysical limits
on photon vacuum birefringence give a bound on 
the coefficients of dimension four
operators of $10^{-32}$~\cite{KM}.

For $n=3$ the constraint from the absence of a strong effect on
energy thresholds involving only electrons and photons is of order
\beq
\eta_3\lesssim (10\, {\rm TeV}/p)^3.
\label{1/p3}
\eeq
Thus we can obtain order unity and even much stronger
constraints from high energy astrophysics, as discussed
shortly.

\subsection{Summary of constraints on LV in QED at $O(E/M)$ }

Since we do not assume universal LV coefficients,
different constraints cannot be combined unless they
involve just the same particle types.  To achieve the
strongest combined constraints it is thus preferable to
focus on processes involving a small number of particle
types. It also helps if the particles are very common and
easy to observe.  This selects electron-photon physics,
i.e. QED, as a useful arena.

The current constraints on the three LV parameters at
order $E/M$---one in the photon dispersion relation and
two in the electron dispersion relation---will now be
summarized.  These are equivalent to the parameters in
the dimension five operators (\ref{dim5}) written down by
Myers and Pospelov.

First, the constraint $|\eta_+-\eta_-|<4$ on the difference
between the positive and negative electron helicity parameters was
deduced by Myers and Pospelov~\cite{MP} using a previous
spin-polarized torsion pendulum experiment~\cite{Heckel} that
looked for diurnal changes in resonance frequency.  (They also
determined a numerically stronger constraint using nuclear spins,
however this involves four different LV parameters, one for the
photon, one for the up-down quark doublet, and one each for the
right handed up and down quark singlets. It also requires a model
of nuclear structure.)

\begin{figure}
\center
\includegraphics[height=6cm]{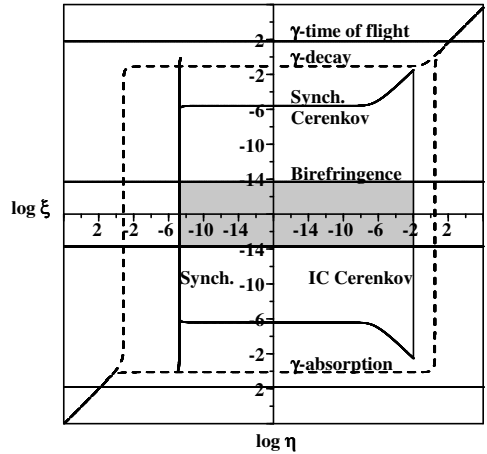}
%\picplace{5cm}{2cm} % Give the correct figure height and width in cm
%
\caption{Constraints on LV in QED at $O(E/M)$ (figure from Ref.~\cite{JLMS}).}
\label{fig:1}       % Give a unique label
\end{figure}

In Fig.~\ref{fig:1} (from Ref.~\cite{JLMS})
constraints on the photon ($\xi$) and
electron ($\eta$) LV parameters are plotted on a
lograrithmic scale to allow the vastly differing
strengths to be simultaneously displayed. For negative
parameters minus the logarithm of the absolute value is
plotted, and a region of width $10^{-18}$ is excised
around each axis. The synchrotron and \v{C}erenkov
constraints are known to apply only for at least one
$\eta_\pm$. The IC and synchrotron \v{C}erenkov lines are
truncated where they cross. Prior photon decay and
absorption constraints are shown in dashed lines since
they do not account for the EFT relations between the LV
parameters.

\paragraph{Vacuum bifrefringence}

The birefringence constraint arises from the fact that the LV
parameters for left and right circular polarized photons are
opposite (\ref{QEDdisp}).  The phase velocity thus depends on both
the wavevector and the helicity. Linear polarization is therefore
rotated through an energy dependent angle as a signal propagates,
which depolarizes any initially linearly polarized signal.  Hence
the observation of linearly polarized radiation coming from far
away can constrain the magnitude of the LV parameter. This effect
has been used to constrain LV in the dimension three
(Chern-Simons)~\cite{CFJ}, four~\cite{KM} and
five~\cite{GK,JLMS,Mitro} terms. The constraint shown in the
figure
derives from the recent report~\cite{CB03} of a high degree of
polarization of MeV photons from GRB021206. The data analysis has
been questioned~\cite{RF03} and defended~\cite{CB03b}, so we shall
have to wait and see if it is confirmed. The next best constraint
on the dimension five term is $|\xi|\lesssim 2\times10^{-4}$, and
was deduced by Gleiser and Kozameh~\cite{GK} using UV light from
distant galaxies.

\paragraph{Photon time of flight}

The $\gamma$ time of flight constraint arises from an
energy dependent dispersion in the arrival time at Earth
for photons originating in a distant
event~\cite{Pavlopoulos,GAC-Nat}, 
which was previously exploited for
constraints~\cite{Schaefer,Biller,Kaaret}. The dispersion
of the two polarizations is larger since the difference
in group velocity is then $2|\xi|p/M$ rather than
$\xi(p_2-p_1)/M$, but the time of flight constraint
remains many orders of magnitude weaker than the
birefringence one from polarization rotation. In
Fig.~\ref{fig:1} we use the EFT improvement of the
constraint of~\cite{Biller} which yields $|\xi|<63$.

\paragraph{Vacuum \v{C}erenkov effect, inverse Compton electrons}

In the presence of LV the process of vacuum \v{C}erenkov
radiation $e\rightarrow e\gamma$ can occur.  The inverse
Compton (IC) \v{C}erenkov constraint uses the electrons
of energy up to 50 TeV inferred via the observation of 50
TeV gamma rays from the Crab nebula which are explained
by IC scattering.  Since the vacuum \v{C}erenkov rate is
orders of magnitude higher than the IC scattering rate,
that process must not occur for these
electrons~\cite{CGlong,Jacobson:2002hd}.  The threshold for vacuum
\v{C}erenkov radiation depends in general on both $\xi$
and $\eta$, however in part of the parameter plane the
threshold occurs with emission of a soft photon, so $\xi$
is irrelevant. This produces the vertical IC \v{C}erenkov
line in Fig. \ref{fig:1}.  One can see from (\ref{1/p3})
that this yields a constraint on $\eta$ of order
$(10~{\rm TeV}/50~{\rm TeV})^3\sim 10^{-2}$.  It could be
that only one electron helicity produces the IC photons
and the other loses energy by vacuum \v{C}erenkov
radiation. Hence we can infer only that at least one of
$\eta_+$ and $\eta_-$ satisfies the bound.

\paragraph{Crab synchrotron emission}

A complementary constraint was derived in~\cite{Crab} by
making use of the very high energy electrons that produce
the highest frequency synchrotron radiation in the Crab
nebula.  For negative values of $\eta$ the electron has a
maximal group velocity less than the speed of light,
hence there is a maximal synchrotron frequency that can
be produced regardless of the electron
energy~\cite{Crab}. Observations of the Crab nebula
reveal synchrotron radiation at least out to 100 MeV
(requiring electrons of energy 1500 TeV in the Lorentz
invariant case), which implies that at least one of the
two parameters $\eta_\pm$ must be greater than
$-7\times10^{-8}$ (this constraint is independent of the
value of $\xi$).  We cannot constrain both $\eta$
parameters in this way since it could be that all the
Crab synchrotron radiation is produced by electrons of
one helicity.  Hence for the rest of this discussion let
$\eta$ stand for whichever of the two $\eta$'s satisfies
the synchrotron constraint.

{\it This must be the same $\eta$ as satisfies the IC
  \v{C}erenkov constraint discussed above}, since
otherwise the energy of these synchrotron electrons would
be below 50 TeV rather than the Lorentz invariant value
of 1500 TeV.  The Crab spectrum is well accounted for
with a single population of electrons responsible for
both the synchrotron radiation and the IC $\g$-rays. If
there were enough extra electrons to produce the observed
synchrotron flux with thirty times less energy per
electron, then the electrons of the other helicity which
would be producing the IC $\g$-rays would be too
numerous~\cite{JLMS}.
It is
important that the {\it same} $\eta$,
i.e.\  either $\eta_+$ or $\eta_-$, satisfies both the
synchrotron and the IC \v{C}erenkov constraints.
Otherwise, both constraints could have been satisfied by
having one of these two parameters arbitrarily large and
negative, and the other arbitrarily large and positive.

\paragraph{Vacuum \v{C}erenkov effect, synchrotron electrons}

The existence of these synchrotron producing electrons can be
exploited to improve on the vacuum \v{C}erenkov constraint. For a
given $\eta$ satisfying the synchrotron bound, some
definite electron energy $E_{\rm synch}(\eta)$ must be present to
produce the observed synchrotron radiation. (This is higher for
negative $\eta$ and lower for positive $\eta$ than the Lorentz
invariant value~\cite{Crab}.) Values of $|\xi|$ for which the
vacuum \v{C}erenkov threshold is lower than $E_{\rm synch}(\eta)$
for either photon helicity can therefore be
excluded~\cite{JLMS}. %t30 added ref to us here.
(This is always a hard photon threshold, since the soft photon
threshold occurs when the electron group velocity reaches the low
energy speed of light, whereas the velocity required to produce
any finite synchrotron frequency is smaller than this.) For
negative $\eta$, the \v{C}erenkov process occurs only when
$\xi<\eta$~\cite{Jacobson:2002hd,KonMaj}, so the excluded
parameters lie in the region $|\xi|>-\eta$.

\paragraph{Photon decay and photon absorption}

Previously obtained constraints from photon decay
$\g\rightarrow e^+e^-$ and absorption $\g\g\rightarrow
e^+e^-$ must be re-analyzed to take into account the
different dispersion for the two photon helicities, and
the different parameters for the two electron helicities,
but there is a further complication: both these processes
involve positrons in addition to electrons.  Previous
constraint derivations have assumed that these have the
same dispersion, but that need not be the
case~\cite{Leh}.  As discussed above, for the $O(E/M)$
corrections this is indeed not so~\cite{JLMS}. Taking
into account the above factors could not significantly
improve the strength of the constraints (which is mainly
determined by the energy of the photons). We indicate
here only what the helicity dependence of the photon
dispersion implies, thus neglecting the important
role of differing parameters for electrons, positrons and their
helicity states.

The strongest limit on photon decay came from the highest
energy photons known to propagate, which at the moment
are the 50 TeV photons observed from the Crab
nebula~\cite{Jacobson:2002hd,KonMaj}.  Since their
helicity is not measured, only those values of $|\xi|$
for which {\it both} helicities decay could be ruled out.
The photon absorption constraint came from the fact that
LV can shift the standard QED threshold for annihilation
of multi-TeV $\g$-rays from nearby blazars such as Mkn
501 with the ambient infrared extragalactic
photons~\cite{Kluzniak,GAC-Pir,SG,Jacobson:2002hd,KonMaj,Comment,S03}.
LV depresses the rate of
absorption of one photon helicity and increases it for
the other.  Although the polarization of the $\g$-rays is
not measured, the possibility that one of the
polarizations is essentially unabsorbed appears to be
ruled out by the observations which show the predicted
attenuation~\cite{S03}.

\subsection{Constraints at $O(E^2/M^2)$?}

As previously mentioned, CPT symmetry alone could exclude the
dimension five LV operators that give $O(E/M)$ modifications to
particle dispersion relation, and in any case the constraints on
those have become nearly definitive. Hence it is of interest to
ask about the dimension five and six operators that give
$O(E^2/M^2)$ corrections. We close with a brief discussion of the
constraints that might be possible on those, i.e. constraints at
$O(E^2/M^2)$.

As discussed above, the strength of constraints
can be estimated by the requirement
$\eta_4 p^4/M^2 \lesssim m^2$, which yields
\beq
\eta_4\lesssim
\left(\sqrt{\frac{m}{1\, {\rm eV}}}\frac{100\, {\rm TeV}}{p}\right)^4.
\label{eta4bound}
\eeq
Thus for electrons, an energy around $10^{17}$ eV is
needed and we are probably not going to see any effects
directly from such electrons. For protons an energy $\sim
10^{18}$ eV is needed. This is well below the UHE cosmic
ray energy cutoff, hence if and when Auger~\cite{Auger}f
confirms the
identity of UHE cosmic rays as protons at the GZK cutoff,
we will obtain a constraint of order $\eta_4\lesssim
10^{-5}$ from the absence of vacuum \v{C}erenkov
radiation for $10^{20}$ eV protons!  Also, from the fact
that the GZK threshold is not shifted, we will obtain a
constraint of order $\eta_4\gtrsim -10^{-2}$, assuming
equal $\eta_4$ values for proton and pion.

Impressive constraints might also be obtained from the absence of
neutrino vacuum \v{C}erenkov radiation: putting in 1 eV for the
mass in (\ref{eta4bound}) yields an order unity constraint from
100 TeV neutrinos, but only if the \v{C}erenkov {\it rate} is high
enough. The rate will be low, since it proceeds only via the
non-local charge structure of the neutrino. Recent
calculations~\cite{Dave?} have shown that the rate is not high
enough at that energy. However, for $10^{20}$ eV UHE neutrinos,
which may be observed by the EUSO and OWL planned satellite
observatories, the emission rate will be high enough to derive a
strong constraint.  The exact value depends on the emission rate,
which has not yet been computed.  For a {\it gravitational}
\v{C}erenkov reaction, the rate
(which is lower but easier to compute than the electromagnetic rate) 
would be high enough for a neutrino from a distant source
provided $\eta_4\gtrsim 10^{-2}$. Hence in this
case one might obtain a constraint of order $\eta_4\lesssim
10^{-2}$, or stronger in the electromagnetic case.

A time of flight constraint at order $(E/M)^2$ might be
possible~\cite{GACsecgen} if gamma ray bursts produce UHE
($\sim 10^{19}$ eV) neutrinos, as some models predict,
via limits on time of arrival differences of such UHE
neutrinos vs. soft photons (or gravitational) waves.
Another possibility is to obtain a vacuum birefringence
constraint with higher energy photons~\cite{Mitro}
(although such a constraint would
be less powerful since the parameters for opposite
polarizations need not be opposite at order
$(E/M)^2$).
If future GRB's are found to be polarized at $\sim 100$ MeV,
that could provide a birefringence constraint
$|\xi_{4+}-\xi_{4-}|\lesssim 1$.

\section{Conclusion}

At present there are only hints, but no compelling evidence for
Lorentz violation from
quantum gravity. Moreover, even if LV is present,
the use of EFT for its low energy parametrization is not necessarily valid.
Nevertheless, we believe that
 the constraints derived from the simple ideas discussed
here are still important.  They allow tremendous
advances in observational reach to be applied in a straightforward
manner to limit reasonable possibilities that
might arise from fundamental Planck scale physics.
Such guidance is especially welcome
for the field of quantum
gravity, which until the past few years has had little connection
with observed phenomena.

%----------------------------

%\printindex
\end{document}